\documentclass[12pt,epsf]{article}
\usepackage{aaspp4}
\usepackage{latexsym}
\usepackage{flushrt}

\begin{document}

%============================================================

\title{The Evolution of Blue Stragglers Formed Via Stellar Collisions.}

\author{J.~A.~Ouellette and C.~J.~Pritchet}
\affil{\vskip 0.2 cm Department of Physics \& Astronomy,
\\University of Victoria,
\\Box 3055, Victoria BC, V8W 3P6 Canada
\\Electronic mail: ouellet@uvastro.phys.uvic.ca, pritchet@uvastro.phys.uvic.ca}
\authoraddr{University of Victoria, Department of Physics \& Astronomy Box
  3055, Victoria BC, V8W 3P6 Canada}
\authoremail{ouellet@uvastro.phys.uvic.ca}

\authoremail{pritchet@uvastro.phys.uvic.ca}

\abstract{

%No stars were mistreated during the writing of this paper....

We have used the results of recent smoothed particle hydrodynamic
simulations of colliding stars to create models appropriate for input
into a stellar evolution code.  In evolving these models, we find that
little or no surface convection occurs, precluding angular momentum
loss via  a magnetically-driven stellar wind as a viable mechanism for slowing
rapidly rotating blue stragglers which have been formed by collisions.
Angular momentum transfer to either a circumstellar disk (possibly
collisional ejecta) or a nearby companion are plausible mechanisms for
explaining the observed low rotation velocities of blue stragglers

Under the assumption that the blue stragglers seen in NGC 6397 and 47
Tuc have been created solely by collisions, we find that the majority
of blue
stragglers cannot have been highly mixed by convection or meridional
circulation currents at anytime during their evolution.  Also, on the
basis of the agreement between the predictions of our non-rotating
models and the observed blue straggler distribution, the evolution of
blue stragglers is apparently not dominated by the effects of rotation.

}

%==============================================================

\newpage
\section{Introduction}

Blue stragglers lie roughly along an extension of a star cluster's main
sequence (MS) and are generally bluer and brighter than the turn-off
(TO) stars.  While several possible mechanisms for their
formation have been proposed (e.g. delayed star formation, binary mass
transfer, binary coalescence, stellar collisions; see Livio, 1993, and Stryker, 1993,
for reviews), 
\nocite{livio93}\nocite{stryker93}
the collision scenario, in which initially unbound stars come into
contact and merge during a dynamical encounter, has received the most
attention recently (e.g. Sandquist \hbox{\em et al.} 1997; 
Ouellette \& Pritchet 1996; Sills, Bailyn \& Demarque 1995).
\nocite{sbh97}\nocite{op96}\nocite{sbd95}
Blue stragglers formed by collisions are of particular
interest because their formation rate tells us about the stellar
interaction rate in their cluster environment and the dynamical
history of the cluster itself (Bailyn, 1995).\nocite{bailyn95}

There are at least two ways to investigate the formation rate of blue
stragglers by collisions.  One is to perform N-body simulations,
modelling the stellar environment and determining the collision rate
directly.  
 This requires knowledge of the stellar interaction
cross-section, local stellar density, mass function, and binary
fraction.  The second 
is to model the 
structure of the merger remnants and to evolve them using a stellar
evolution code.  Just as  the age of a cluster is found by
comparing theoretical 
isochrones, based on standard stellar models,  with an observed
colour-magnitude diagram (CMD), comparison of the evolution of the
merger models to the observed numbers and distribution of blue
stragglers in the CMD can tell
us about the distribution of lifetimes, and the production rates, of
these stars.

In this paper, we develop and evolve models of collisional merger
remnants and compare the predictions of these models 
with the observed blue straggler distributions in NGC 6397 and 47 Tuc.

%=================================================================

\section{Predictions of the Collision Scenario --- A Review.}

There is direct and indirect evidence that blue stragglers
are more massive than the turn-off (TO) stars in their parent clusters
(spectroscopic measurements:
Shara \hbox{\em et~al.} 1997,\nocite{shara97}
Rodgers \& Roberts 1995\nocite{rodgers95};
mass segregation: e.g. Sarajedini \& Forrester 1995\nocite{sf95},
Lauzeral \hbox{\em et al.} 1992\nocite{lauzeral92},
Mathieu \& Latham 1986\nocite{ml86}),
suggesting  a formation mechanism which involves combining the mass
of at least two stars into one massive star
(e.g. Livio 1993\nocite{livio93}).
Although there
are a few possible mechanisms by which this can be achieved,
 this paper is concerned with
only one of them --- the collision scenario. This
requires that two stars come into direct contact during a strong
dynamical encounter and then merge, leaving behind a massive remnant.  As
well as being more massive, the collisionally merged star will have
other properties which we will summarize here.

Hills and Day (1976)\nocite{hnd76} 
investigated the possibility that collisions could
have occurred between single stars in globular clusters (GCs) within
the lifetimes of the clusters.  They
found that, in the dense cores of some GCs, the timescale for single
star interactions is short enough that a significant number of
collisions would have occurred.  Also, because the impact parameter at
the time of the collision is random, the collisions will,
on average, occur off-axis.  Angular momentum conservation then requires
that the merged star be rapidly rotating.  Hills \& Day hypothesised
that the excess energy from the collision would mix nuclear
processed material throughout the merger remnant, as well as mixing
hydrogen into the core, 
essentially ``resetting the nuclear 
clock'' to the chemically homogeneous zero-age main sequence (ZAMS)
state.

Hoffer (1983)\nocite{hoffer83}
 and Leonard (1989)\nocite{leonard89} 
extended the analysis of collision
probabilities to include binary-single star and binary-binary
interactions.  Because of the increased physical cross-section
(essentially the semi-major axis of the binary) and the additional
gravitational focussing due to the larger mass of the bound pair, the
probability that a collision will occur is greatly increased if one or
more of the interacting objects is a binary.  Since the rate at which
binary interactions will occur in a population of stars will
increase with the binary fraction, we  expect to see a
significant collision rate in  clusters with a large binary fraction
and in the cores of clusters, where most of the  binaries are expected
to be found due to mass segregation.

Leonard and Fahlman (1991)\nocite{lnf91} 
found that the rate of production of blue
stragglers in NGC 5053 needed to be at least 1 blue straggler every ${\sim}
{2.5\times}10^8$years, assuming that the average lifetime of a blue
straggler is ${\sim}{6\times}10^9$years, to maintain the observed
numbers.  From their scattering experiments, a strong binary-binary
interaction should occur every ${\sim}{2.2\times}{10^7}$years, with only
1 in 20-40 such interactions actually resulting in a stellar collision
(hence, one blue straggler produced every
${\sim}{({4}-{8})\times}{10^8}$ years). 
  According to Sigurdsson \&
Phinney (1993), the timescale for exchanges during strong interactions
is significantly shorter than the timescale for collisions.  Also,
such exchanges tend to produce a net hardening of the 
binary, which will lengthen the timescale for collisions.

Leonard (1996)\nocite{leonard96} 
noted that the high frequency of binaries observed
among blue stragglers (e.g. 
Kaluzny \hbox{\em et al.}\ 1997\nocite{kaluzny97b},
Kaluzny 1997\nocite{kaluzny97a},
Edmonds \hbox{\em et al.}\ 1996\nocite{edmonds96})
could be accounted for by binary-binary
interactions.  A dynamical encounter that is strong enough to result
in a direct collision between two of the component stars is also
likely to result in a third star being captured into an eccentric orbit
around the newly formed blue straggler; the fourth star is ejected
from the system, removing the excess binding energy.

These dynamical studies treated the interacting stars as point
sources, and assumed that a direct collision would take place
and form a blue straggler if two of the binary components approached
within some factor of their radii, ignoring the 
dynamics of the stellar collision itself.    Benz
\& Hills (1987, 1992)\nocite{bh87}\nocite{bh92} 
studied stellar collisions by performing
smoothed particle hydrodynamic (SPH) simulations of colliding $n=3/2$
polytropes.  They found that the resulting merger remnant's properties
depended on the relative mass of the parent stars.  During collisions
involving equal mass stars, Benz \& Hills found that the material from
the parent stars was distributed throughout the merger remnant,
resulting in a star with a roughly homogeneous chemical composition:
essentially a ZAMS star.
Their collisions involving unequal mass stars resulted in a merger
remnant  with the material from the less evolved, low mass parent star
in the core of the remnant;
the material 
from the high mass star was, in contrast,  mixed throughout.  In
either case, the merger remnant roughly resembled a ZAMS star due
to the increased hydrogen content in the core.

Lombardi, Rasio \& Shapiro (1996, hereafter LRS)\nocite{lrs96} 
found that the high
degree of mixing 
which Benz \& Hills had observed in their merger remnants was largely
an artifact of the low resolution of their simulations and their
choice of $n=3/2$ polytropes as approximations to 
GC stars.  LRS noted that evolved GC stars would be more centrally
condensed than $n=3/2$ polytropes and so chose to use $n=3$ polytropes
to approximate the structure of TO stars, and $n=3/2$ polytropes for
 lower mass stars.  SPH simulations
using these approximations demonstrated that there should be little
or no hydrodynamical mixing during a stellar collision.

Benz \& Hills (1987, 1992) and LRS found that significant mass loss could occur
during grazing collisions; in the high impact velocity collisions of
Benz \& Hills, the two stars could be completely disrupted, leaving no
single remnant behind.  If the star was not disrupted during the
collision, it was swollen to pre-main sequence sizes by the injection
of orbital kinetic energy into the stellar material.
  The greatest
amount of mass loss for a given impact velocity occurred during off-axis 
collisions.

The high rotation velocity which the collision scenario predicts for 
 blue stragglers is one of its potential weak points.  In
open clusters, such as M67 
(Peterson \hbox{\em et al.} 1984\nocite{peterson84}, Mathys
 1991\nocite{mathys91}, Pritchet \& Glaspey 1991\nocite{prit91}),
blue stragglers are not observed to be rapidly rotating.  The only
reported observation of the rotation velocity of a blue straggler in a
globular cluster 
(Shara \hbox{\em et al.} 1997)\nocite{shara97},
 $V\sin i = 155 \pm 55$km/s, is
high, but not necessarily unusual for normal stars of the same
spectral type (A7V: $\overline{V}_{rot}{\sim} 
{130}-{170}$km/s, Lang 1992) nor is it rotating near to its
break-up velocity of $\sim{410}$km/s (estimated using the mass and $\log
g$ from Shara \hbox{\em et al.}). 
\nocite{astrophdata} 
Leonard \& Livio (1995)\nocite{lnl95} 
suggested that the rotation
velocity of a blue straggler could be lowered by angular momentum loss
(AML) via a magnetically driven stellar wind,  similar to the
mechanism proposed for 
pre-main sequence
stars by 
Tout \& Pringle (1992)\nocite{tout92}.  
For this mechanism to work, the blue
straggler would need to be left in a largely convective, Hayashi-like
phase after  the collision.   

In addition to providing an AML mechanism for blue stragglers, a highly
convective phase would also mix the stars, homogenizing them despite
the fact that they will be initially inhomogeneous (LRS).  
Bailyn \& Pinsonneault (1995)\nocite{bp95} 
and Sills, Bailyn \& Demarque (1995)\nocite{sbd95} 
found that such a high degree of mixing is needed to explain the
colours and luminosities of blue stragglers seen in some globular
clusters.   On the other hand, 
Ouellette \& Pritchet (1996)\nocite{op96} 
found that
blue stragglers tend to avoid the ZAMS and that a high degree of
mixing during a pre-main sequence phase is essentially ruled out
for most blue stragglers by their distribution on the
CMD.  If significant surface convection occurs, though not
necessarily enough to mix the
star, abundance anomalies may be observable.  In fact, many of the
blue stragglers in M67 have been observed to be underabundant in
lithium 
(Pritchet \& Glaspey, 1991\nocite{prit91})
 and have anomalous CNO abundances 
(Mathys, 1991)\nocite{mathys91},
 possibly indicative of mixing.

In short, the properties of blue stragglers formed through collisions
are that they should be more massive than the cluster TO stars, they {\em
  might} be rapidly
rotating, they {\em might} have surface chemical anomalies, and they are
likely to have a companion in an eccentric orbit.

%=================================================================

\section{Structure of Merger Remnants.}

Scattering experiments, such as those of 
Leonard (1989)\nocite{leonard89}
 and
Sigurdsson \& Phinney (1993)\nocite{sig93},
 tell us at what rate we can expect
stellar collisions to occur. However, in order to test the accuracy and
viability of the collision scenario, we need to know the timescales
over which collisional merger remnants will be observable as blue
stragglers.  This can 
be done by creating stellar models from the predictions of the SPH
simulations  of stellar collisions, and following their evolution with
a stellar evolution 
code. Once
these timescales are known, they can be combined with the results of
 scattering experiments to predict the number of blue stragglers
and other remnants of such strong interactions 
(Bailyn, 1995\nocite{bailyn95}) 
which should be
%
%add more about the byproducts of collisions rates.
%
observable in a cluster.  Also, calculation of the 
evolution of merger remnants allows us to make predictions concerning the
distribution of blue stragglers in the CMD.  Comparison of this
distribution with observations allows us to infer whether collisional
mergers are likely to have occurred in a population of blue stragglers,
and even the dynamical history of the cluster itself.

As mentioned earlier, LRS have studied stellar collisions by
performing SPH simulations of colliding polytropes.  Their results
showed that little, if any, hydrodynamical mixing occurred during the
collision.  The composition profiles of their merger remnants can be
understood as an `entropy stratification' of the stellar material
during the collision:  the gas from the parent stars settles in  the
merger remnant such that the final entropy profile increases from the
core outwards.   This entropy stratification can result in some
unusual chemical profiles (as shown by LRS), but also allows some
prediction of the chemical profile of a collisional remnant ---  which we
will use in this study.

The collisions studied by LRS were between equal mass polytropes and
unequal mass polytropes, for a variety of different masses.  We choose
here to model only mergers between equal mass stars  and mergers
between a TO star and a lower mass star.   Throughout the rest of this
paper, we refer to these merger events as ``equal mass mergers'' and ``unequal
mass mergers'', respectively.  
Although there are any number of possible combinations of parent star
masses which would result in a particular merger mass between
$1M_{TO}$ and $2M_{TO}$, the mergers considered here represent the
extremes of hydrogen content: equal mass mergers will result in the
highest possible hydrogen content for a particular merger mass,
whereas unequal mass mergers will result in the lowest possible
hydrogen content.

\subsection{Predictions from Entropy Stratification.}

In their simulations, LRS approximated MS TO stars
by $n=3$ polytropes while lower mass stars were approximated by
$n=3/2$ polytropes or composite polytropes.  Their initial polytropic
models had entropy and density profiles similar to those shown in
Figure 1.  During a collision involving a $n=3$ polytrope
(${\sim}0.80M_\odot$, MS TO
star) and a $n=3/2$ polytrope (${M}{\lesssim}{0.40M_\odot}$, lower MS star),
 entropy stratification predicts that the material of the lower mass,
and presumably less evolved, $n=3/2$
polytrope will settle into the core of the merger remnant, bringing
with it a fresh supply of hydrogen.  The subsequent evolution of the
merger remnant will be strongly affected by the amount of hydrogen
brought into the core by the entropy stratification.  This
stratification of the material also provides a simple explanation of
why no nuclear processed 
material is brought to the surface of the merger remnant during  a collision.

If the entropy of the stellar gas is not modified during a collision,
the distribution of the parent stars' material throughout the merger
remnant can be found using the entropy profiles in Figure 1 --- this
leads directly to the merger remnant's chemical profile if those
of the parent stars are also known.
  However, shock heating during
the collision can modify the entropy of the gas depending on the
dynamics of the collision and the form of viscous dissipation chosen
for the SPH (LRS).
 During relatively gentle collisions, such as
the head-on, parabolic collisions studied by LRS, little shock heating
will occur and so the entropy of the gas at the time of the collision can
be used to determine the final merger profile (See section~\ref{sec:shock}).

\subsubsection{Stellar Collisions versus Polytropic Collisions.}

During the previous discussion we have stressed that the
simulations of LRS and Benz \& Hills (1987, 1992) describe {\em polytropic}
collisions.  The reason for this pointed emphasis is that stars,
especially evolved ones in GCs, are not polytropes.  

The structure of a polytrope is given by 
\begin{equation}
P(r) = K\rho(r)^{(1+1/n)}\label{eq:polytrope}
\end{equation}
where $P$ is the pressure, $\rho$ is the gas density, $n$ is the
polytropic index and $K$ is a constant.  Using this relationship as a
constraint, $\rho(r)$ and $K$ can be found using only two
of the equations of stellar structure: the
equations of hydrostatic equilibrium and mass continuity.  For a
simple polytrope, the molecular weight $\mu$ is constant throughout,
 in which case both $K$ and $n$ are also constants.  This allows
the structure of the polytrope to be determined with no information
about sources of opacity or nuclear processes.  

The entropy profile of a polytrope can be found by assuming an ideal
gas equation of state
of the form 
\begin{equation}
P = A\rho^{\Gamma_1}, \label{eq:idealeos}
\end{equation}
 where $\Gamma_1$ is one of the adiabatic exponents and
is defined by 
\begin{equation}
{{dP}\over P} + \Gamma_1 {{dV}\over V} = 0. \label{eq:gamma1}
\end{equation}
If radiation pressure within the gas can be ignored, then $\Gamma_1 =
\gamma = c_P/c_V = 5/3$ for an ideal, monatomic gas.  Equating the pressure
from the polytropic 
structure to the pressure from the equation of state yields $A$
throughout the gas.  The relationship between the entropy $S$ and the
constant $A$ can be found by integrating the first law of
thermodynamics, $TdS = dU + PdV$, using the definition of $\gamma$ and
Equation~\ref{eq:gamma1}.
This yields 
\begin{equation}
S = {{N_0k}\over{\mu\gamma}} {\rm ln}A + const \label{eq:entropy}
\end{equation}
where $N_0$ is Avagadro's number
 and $k$ is Boltzmann's constant.  From the
form of the polytropic structure equation (Eq.~\ref{eq:polytrope}) and the equation of
state (Eq.~\ref{eq:idealeos}),
if $1 + 1/n$ is not equal to $\gamma$, then $A$, and hence
the entropy $S$, will not be constant throughout the star.  Polytropes
with $n=3/2$ will have $1 + 1/n = \gamma$ and so will have
constant entropy.

Polytropes are reasonable approximations to ZAMS stars since they,
like polytropes, are chemically homogeneous, or at least nearly so.
As pointed out by LRS, $n=3$ polytropes approximate the structure of
radiative stars while $n=3/2$ polytropes have a structure similar to
that of fully convective stars.  However, although GC TO stars are
radiative, they are far from being chemically homogeneous.  Figure 2
shows the entropy and density profiles for a GC TO star and lower MS
stars for a cluster like 47 Tuc.  The differences between the entropy
and density profiles of GC stars and the polytropes shown in Figure 1
are largely due to the chemical inhomogeneity of the GC stars.  The
increased molecular weight toward the centre of an evolved star
results in a lower pressure at a fixed density and temperature,
requiring the star to adjust its internal structure to
maintain hydrostatic equilibrium.  Equation~\ref{eq:entropy} shows that this increase in 
the molecular weight results in a decrease in the entropy of the gas.  

The decrease of the entropy in the core of an evolved star will
directly affect the final chemical profile of a merger found through
entropy stratification.  From Figure 2, it is obvious that no material
from a low mass star will be able to penetrate to the core of a MS TO
star, leaving the merger remnant with a helium-rich core, unlike the
collision between two equivalent polytropes.  The evolution of two
such merger remnants would be completely different if no mixing takes
place in a subsequent convective phase of evolution.

Comparison of Figures 1 and 2 shows that an $n=3$ polytrope
underestimates the central density of a TO star by more than a factor of
ten.  This in itself is enough to argue that no hydrogen-rich material
from a lower mass star will penetrate to the core of a merger
involving a TO star and a lower MS star (an unequal mass merger in our
terminology).   Entropy stratification of
the material during such a collision will produce a merger
remnant with a dense, helium-rich core.

A collision involving equal mass stars or one involving equal mass
polytropes should produce a merger remnant with a composition profile
nearly identical to those of the parent stars.

\subsubsection{Shock Heating.}
\label {sec:shock}

As mentioned earlier, shock heating can modify the entropy of the gas
during the collision and so affect the distribution of material
throughout the merger remnant.  If the amount of shock heating is
significant, the structure of the merger remnant could differ considerably
from what would be expected if the unmodified entropy profiles of the
parent stars were used to estimate the final structure.

The amount of shock heating due to the passage of a shock can be
estimated from the pre-shock state of the gas, and the velocity of the
shock. Making the simple assumptions that the sound velocity within
the gas is equal to the adiabatic sound speed, $c_s^2 = {{\partial
    P}\over{\partial\rho}}$, and that there is no change of state due
to the passage of the shock (i.e. $\gamma_{final} =
\gamma_{initial}$) then, taking the Mach number of the shock as
$M_0 (={V_{S}\over c_s})$, the change in entropy after the
passage of the shock is (e.g. Shore 1992)\nocite{shore92}
\begin{equation}
\label{eq:shock}
\Delta S = S_1 - S_0 = {N_0k\over{\mu \gamma}}
\ln\Bigg[\Bigg({P_1\over
  P_0}\Bigg)\Bigg({\rho_1\over\rho_0}\Bigg)^{-\gamma}\Bigg]
\end{equation}
where
$${\rho_1\over\rho_0} = {{(\gamma + 1) M^2_0}\over{(\gamma -
    1)M^2_0 + 2}}$$
and
$${P_1\over P_0} = {{2\gamma  M^2_0 - (\gamma - 1)}\over{\gamma + 1}}$$
Here, the subscript 0 refers to the conditions on the immediate
pre-shock side of the shock discontinuity, and the subscript 1 refers
to those conditions immediately on the post-shock side of the
discontinuity.  Figure 3 
shows the effect of shock heating on the entropy profiles of the stars
shown in Figure 2 under the assumption that the shock velocity remains
constant throughout the star.$\!\!\!$
\footnote{This also neglects the effects of viscous dissipation, which would
  affect both $\Delta S$ and the velocity of the shock.  The initial
  shock velocity is taken to be the impact velocity (See Sec. 4).
  Typically, the shock has a Mach number $\sim 5$ at
  $M(r)\sim0.99M_{*}$, and $\lesssim 2$ 
  at the centre. 
}
It is obvious that there is potentially significant shock heating near
the surface of the stars and very little in the cores.  In particular,
entropy stratification of the parent stars' material using the
`post-shock' entropy profiles in Figure 3 would produce a very similar
stratification to that found using the pre-shock profiles.  The
change in entropy near the surface of the star may result in some
mixing of the surface layers, however, since these layers should
contain no nuclear processed material, this would not significantly affect the
subsequent evolution of the merger remnant.

The amount of shock heating calculated above is undoubtably an
underestimate --- an actual collision would result in multiple shocks,
especially during grazing collisions, and would have a much more complex
geometry than that used in the above calculation.  A more accurate
estimate of the amount of shock heating could be found through
simulations using SPH, which has been shown to model
shocks well, depending on the form of artificial viscosity used
(Monaghan 1992)\nocite{monaghan92}.  Despite the rough nature of this
estimate of the amount of shock heating, the {\em relative} amounts of
shock heating should be correct: that is, the envelopes of the stars
are more affected by the shocks than the interiors, and that the amount
of shock heating is not enough to greatly affect the stratification of
nuclear processed material when dealing with dissimilar stars.

\subsection{Physical Structure of Merger Remnants.}

Leonard \& Livio (1995) have suggested that a blue straggler formed through a
stellar collision would resemble a pre-main sequence star immediately
after the collision.  The SPH
simulations of LRS show that the remnant of a collision involving two
polytropes can be several times larger than either of the parent
stars;  however, unlike a pre-main sequence star, the central density of the
remnant will be much higher than that found on the
Hayashi track.  The decrease in central density of polytropic
mergers relative to the parent stars is generally less than $50\%$,
and is as little as ${\sim}10\%$ in head-on mergers (from the
simulations of LRS).  Because of the
size and dense core of a merger remnant, it will more closely resemble
a red giant branch star than a star on the Hayashi track.

%==============================================================

\section{Construction of Initial Models.}

Despite the facts that the SPH simulations of LRS use polytropes rather
than evolved stars, and that the differences between the two species of
objects can be
extreme, we can use their results as the basis for producing  models
for input into a stellar evolution code.  Entropy stratification of
the material, which is a direct consequence of the dynamics of the
fluid interaction and a requirement for the SPH fluid stability (see
LRS for a discussion), can be used to predict the chemical profile
of the merger remnant.  Although there is no equivalent procedure for
predicting the density profile of the remnant, the simulations of LRS
show that the 
central density will not, in general, be much lower than that of the
parent stars --- this can be used as an additional constraint when
constructing models.  

Since head-on, parabolic collisions are relatively gentle, they are
also the easiest to approximate for our purposes.  We choose to ignore
mass loss during the collision, the effects of rotation and the
departure from spherical symmetry it produces.  Because head-on
collisions produce the least amount of mass loss and lowest rotation
rates, these approximations are reasonably valid.  We also choose to
ignore shock heating during the collision and use the pre-collision
entropy profiles of our parent stars to define the chemical profile of the
merger product using entropy stratification.  The amount of shock
heating during a head-on 
collision is small (Sec~\ref{sec:shock} and LRS) and so should not
strongly affect the end 
result. 

The initial models we used to investigate blue straggler evolution
were formed in three steps from a series of standard stellar models of
the appropriate age and metallicity.  First, for collisions between a TO star
and a lower MS star, an appropriate TO model was scaled in mass so
that its mass agreed with the total mass of the two parent stars; for
equal mass mergers, the mass of one of the parent stars  was scaled.
Next, the scaled model's composition profile was 
replaced with the chemical profile found by entropy stratification of
the parent stars.  This model was then expanded to simulate the
expansion of the star during the collision.

The expansion of the model was performed by adding an additional term
$\epsilon_x$ to the energy balance equation of stellar structure:
$${\partial{L}\over\partial{M}} = \epsilon_{nuc} - \epsilon_\nu +
\epsilon_{grav} + \epsilon_x,$$
where $\epsilon_{nuc}$ is the nuclear energy generation rate,
$\epsilon_\nu$ is the energy loss due to neutrinos, and
$\epsilon_{grav}$ is the energy generation due to contraction of the
star ($=-T{\partial{S}\over{\partial{t}}}$).
If $\epsilon_x$ were held constant throughout the star, the central
density would decrease rapidly as the star expanded and the star would
end up on, or near,
the Hayashi track (this is similar to the procedure which 
Sandquist \hbox{\em et al.} (1997)\nocite{sbh97} 
used to produce their Hayashi models).  As
mentioned earlier, the density of the core does not decrease by a
large factor during the collision; to ensure this, we assumed that $\epsilon_x$
falls off as $1/\rho$ (see below).  During the expansion of our models, the
central density does not decrease by more than $20\%$ which, as
explained below, results in our final models being virtually
independent of the form of $\epsilon_x$ after they have relaxed to the MS.

One additional consideration is how much energy to inject into the
star before allowing it to contract to the MS and evolve normally.  We
have chosen to use a criterion which takes into account the binding
energy of the parent stars and the kinetic energy of the collision.
The binding energy of a star can be expressed as
$$E_{bind} = - q{GM^2\over R},$$
where $M$ and $R$ are the mass and radius of the star, and $q$ is
related to the degree of central concentration 
(Cox \& Giuli, 1968)\nocite{cng68}:
$$ q = \int^1_0 {M(r/R)/M\over{r/R}}\cdot{{\rm d}M(r/R)\over M}.$$
In the centre of mass frame of the two stars the orbital energy is
$$E_{orb} = {1\over 2}\mu v^2 - {GM\mu \over r},$$
which, for a parabolic orbit, is conveniently equal to zero. (Here, $M$
is the total mass of the two stars, $M_1 + M_2$, and $\mu$ is the
reduced mass of the system, $\mu = {{M_1}{M_2}\over{M_1 + M_2}}$.)  Hence,
the kinetic energy of the orbit is
$$K = {GM\mu \over r}.$$
Assuming that one half of the orbital kinetic energy goes into
expanding the merger remnant, the final binding energy of the merger
remnant is $E^\prime_{bind} = E_{bind_1} +  E_{bind_2} + {1\over 2}K$.  For
the purposes of determining the kinetic energy at the time of the
collision, we have set the separation of the two stars $r$ equal to
the average of their radii, $r = {R_1 + R_2 \over 2}$.  The expansion
of the model is halted when its binding energy equals
$E^\prime_{bind}$.  For the same reasons that the form of $\epsilon_x$
is not critical, as long as the density constraint is obeyed (discussed below), the structure of the blue straggler
after it has contracted to its MS is not highly dependent on the exact
value of $E^\prime_{bind}$, although the duration of any convective
zones during the pre-main sequence evolution is affected.

One might be concerned about the ramifications for the models of assuming an
energy injection term proportional to $1/\rho$.  We have investigated
this by adopting different energy 
injection schemes and comparing the resultant models.  We find that,
although the tracks that the models follow on the CMD 
as they expand and contract can differ considerably, the final models that satisfy
the above criteria for 
energy and central density are very similar and evolve identically
 after they have contracted to the MS.  This can be explained
using the Vogt-Russell 
Theorem, which states that the structure of a star in hydrostatic and
thermal equilibrium is uniquely determined by the total mass and the run
of chemical composition throughout the star (Vogt 1926\nocite{vogt26},
Russell 1927\nocite{russell27}; see e.g. Cox \& Giuli, 1968).
Thus, if the amount of mixing which takes place is not significantly
affected by the form of the energy injection term, the evolution of
the merger remnant, at least after it has contracted to the MS, should
also be unaffected by the choice.  Our tests using different forms of
$\epsilon_x$ show that this is the case as long as the central density
does not decrease beyond what is indicated by the SPH simulations.

As pointed out by the referee, our energy injection scheme is similar
to that used by Podsiadlowski (1996)\nocite{pod96} in his
investigation of the response of stars to heating by tidal effects.
In his exploratory paper, Podsiadlowski investigated the effect of
various forms of energy injection (e.g. centrally concentrated,
uniform, surface) on a $0.8M_\odot$ GC ZAMS star --- the different forms
of energy injection were intended to approximate the zones in which
tidally excited oscillations might dissipate their energy.
Podsiadlowski found that the response of the star and its structure
after a fixed amount of energy had been injected were strongly
dependent upon the way in which energy was injected.  Our initial
experiments into the various forms of $\epsilon_x$ were quite similar
to Podsiadlowski's and lend support to his findings; the differences
in the final structures from our early experiments and those of
Podsiadlowski's were simply due to the fact that our initial models
were evolved, whereas his were chemically homogeneous.  Unlike
Podsiadlowski's investigation, however, the form of $\epsilon_x$ for
our models is constrained by the results of SPH: the central density
of head-on mergers does not decrease dramatically from that of the
parent stars'. From our experience, which is similar to what is
reported by Podsiadlowski, forms of $\epsilon_x$ which are uniform
throughout the star or are centrally concentrated result in a rapid
decrease in the central density --- by the time enough energy is
injected to meet our energy constraint ($E^\prime_{bind}$ -- discussed
earlier), the star's structure would resemble that of a star on the
Hayashi track.

\subsection{Cluster Parameters.}

We have chosen to compare our models of merger remnants with the blue
stragglers observed in NGC 6397 and 47 Tuc.  Both clusters have high
central densities (see Table 1), a fact that makes the blue stragglers
in these clusters excellent candidates for formation by collisions. 

The stellar evolutionary code used to produce and evolve our merger
models is a modified version of the 
VandenBerg \hbox{\em et al.} (1997)\nocite{vdb97}
stellar evolution code and opacities.
  For each
cluster, a grid of stellar models was produced, and isochrones were
extracted using the method of Bergbusch \& VandenBerg (1992).  Ages
and TO masses were found for each cluster by matching the TO
luminosity with these isochrones; absolute TO luminosities were found
by using the ``best-fit''
$M^{HB}_V$-[Fe/H] relationship of 
Chaboyer \hbox{\em et al.} (1996)\nocite{chaboyer96}
 to obtain a
distance modulus for each cluster.  The ages derived in this manner agree
with those derived by Chaboyer \hbox{\em et al.}, and are shown in
Table 1 along with the other derived cluster parameters.  It should be noted
that, although the ages and TO masses derived here are dependent upon
the $M^{HB}_V$-[Fe/H] relationship or distance calibration chosen, the
final characteristics of our models (other than mass) are only weakly
dependent upon the choice --- in particular, the surface convection
seen in the 
models presented here becomes less effective if the distance scale is
increased.

%==========================================================

\section{Results and Discussion.}

Blue straggler models were produced as described in the previous sections for
both unequal mass and equal mass mergers, and for masses up to twice
the cluster TO mass.  Figures 4 and 5 show the evolution of these
models on the CMD.  As would be expected, because of the chemical inhomogeneity
of the merger remnant due to entropy stratification, the stars tend to
avoid the ZAMS; only the low mass, relatively unevolved, equal mass
mergers approach what would appear to be normal ZAMS stars.  The
differences between the tracks for the mergers of NGC 6397 and 47 Tuc
are largely due to the difference in cluster metallicity.  These differences,
including convection and a comparison with the
observations, will be discussed in the next sections.

\subsection{Surface Convection.}

Leonard \& Livio (1995)\nocite{lnl95}
 and 
Sills, Bailyn \& Demarque (1995)\nocite{sbd95}
 have
suggested that blue stragglers must become largely convective during
their post-collision, pre-main sequence phase of evolution to explain
both the observed rotation 
velocities  and colours.  While the blue straggler
models of 
Sandquist \hbox{\em et al.}\ (1997)\nocite{sbh97}
 lend some support to this, our
models do not, nor do the similar models of  
Sills \hbox{\em et al.}\ (1997).\nocite{slbd97}

Sandquist \hbox{\em et al.}\ (1997) performed SPH simulations of mergers of equal mass
polytropes and evolved the products of these mergers by imposing the
resultant composition profile on to  standard stellar models of the
same mass as the mergers, and then forcing the stars to expand until they
had reached the 
Hayashi track.  Their models developed deep surface convection,
enough to bring at least some helium to the surface of the stars.
However, forcing the stars onto the Hayashi track would 
decrease the central density beyond what is observed to occur in SPH
simulations (LRS) --- thus requiring an additional energy source  which would not be present in the actual
collision.  In addition, the deep convective envelopes seen by
Sandquist \hbox{\em et al.}\ are a consequence of the fact that their
models are 
forced on to the Hayashi track, at which point the low surface
temperatures will 
require surface convection.  Surface convection will persist until
the opacity in the outer regions of the star decreases to the point
where radiative transfer becomes efficient ---  either when the star has
evolved to higher 
surface temperatures, or  convection has brought a significant
amount of helium to the surface.

Sills \hbox{\em et al.}\ (1997) created initial blue straggler models directly
from the entropy and density profiles produced by SPH. None of their
models developed any surface convection until they had evolved onto
the red giant branch.  The 
total lack of surface convection during the pre-main sequence phase
prevents any helium-rich material from being dredged up to the surface
layers of the star, and makes spin-down
of a rapidly rotating blue straggler by a magnetic wind mechanism
(Leonard \& Livio, 1995) implausible.  However, because
the outer portions of SPH merger remnants are not necessarily in
dynamical equilibrium, Sills \hbox{\em et al.}\ found it
necessary to extrapolate the outer structure of their merger models
from the purely radiative interior, forcing the envelope to be
radiative, at least initially.

Surface convection does occur in some of our models during the
pre-main sequence phase, unlike the models of Sills \hbox{\em et al.}, but not to
the extent found by Sandquist \hbox{\em et al}..  As in standard
models evolving to the MS, the depth and duration of the surface
convection zones in our models depends on the star's mass.  The most
massive merger 
remnants (${M}{\sim}{1.8-2.0}\times{M_{TO}}$), whether formed by an unequal mass or
equal mass merger, never develop surface convection zones during the
phases prior to reaching the MS.  The
convection zones seen in our lower mass models are generally deeper and
longer lived in the lowest mass stars, decreasing in depth and
duration for mergers of higher mass.  The convective zones typically
contain ${\lesssim}{4\%}$ of the star's mass and usually last for less than a
few times $10^6$years.

\subsubsection{Consequences of Surface Convection and Angular Momentum
Loss.}

The thin, short-lived surface convection  seen in our models, and the
absence of surface convection in the models of Sills \hbox{\em et al.}\ (1997),
precludes any 
wind-driven AML mechanism for slowing down the rapidly rotating blue
stragglers predicted by the collision scenario.  However, blue
stragglers are not observed to be rapidly rotating in open clusters
where, despite the comparatively low stellar density, it is 
possible for a fraction of blue stragglers in open
clusters to be formed by collisions 
(Leonard 1996,\nocite{leonard96}
 Leonard \& Linnell 1992).\nocite{lnl92}
 To account for the low rotation velocities, there must be  either  an additional AML mechanism acting which is
not dependent upon surface convection, or the blue stragglers in open
clusters are being formed by a mechanism other than
collisions which might produce more slowly rotating blue stragglers (e.g. binary 
coalescence, binary mass transfer), in which case the estimated numbers of
collisionally generated blue stragglers in open clusters is incorrect.
Even if the production of blue stragglers by collisions is ruled out in
open clusters other formation mechanisms can be assisted, or
accelerated, by strong dynamical interactions (Sigurdsson \& Phinney,
1993)\nocite{sig93}.

Little is known about the rotation of blue stragglers in GCs, although
the stage has been set to rectify this problem by the observations of
Shara \hbox{\em et al.}\ (1997).\nocite{shara97}
  BSS-19 (Paresce \hbox{\em et al.}, 1991)\nocite{paresce91}
 in 47 Tuc is
estimated to have a mass of $1.70\pm0.40M_\odot,\!\!\!$
\footnote{Comparing the observations of effective temperature
  ($T_{eff}=7630\pm 300$K) and surface gravity ($ \log g = 4.09 \pm
  0.1$) directly to our models yields a mass estimate of $1.55\pm0.10M_\odot$,
  independent of distance, in reasonable agreement with the determination of
  Shara \hbox{\em et al.}.}
 compared to the
cluster TO mass of ${\sim}{0.86M_\odot}$, and a high rotation velocity ($V\sin i
= 155 \pm 55{\rm km/s}$).  From our models, this star should not
have had a convective envelope at any time during its ``pre-blue
straggler'' phase, assuming it was created by a collisional merger,
and so should contain the same angular momentum with which it was
created,  unless an AML mechanism other than a
magnetically driven wind has acted upon it.  During its contraction 
to the MS, its rotational velocity should have increased
from its initial value
by a factor of ${\sim}{5-6}$, due to angular momentum conservation, implying
an initial rotation velocity  of
$\sim30$km/s.  From the
results of LRS, this would have required a nearly head-on collision
which, although not unlikely, is less probable than an off-axis
collision.  Hence,  it is more probable that BSS-19, if created by a
collision,  is either inclined
close to the line of sight ($\sin i\lesssim{\pi\over4}$), has
experienced some AML, or was instead formed through some
mechanism other than a collision, as suggested by Shara \hbox{\em et al.}.

It is possible that AML from an initially rapidly rotating blue
straggler could be achieved by angular momentum transfer to a
circumstellar disk, possibly ejecta from the collision, or to a nearby
companion, possibly captured during a binary interaction.  
Cameron \hbox{\em et al.}
(1995) found that angular momentum transfer to a circumstellar disk is
an extremely efficient mechanism for slowing the rotation of stars as
they contract to the main sequence. This mechanism does require that
the star have a convective envelope for the generation of a magnetic
field, but the field strength required to shed a given amount of
angular momentum is not necessarily as high as
that needed for a wind-driven mechanism.  
Angular momentum transfer to a nearby companion has the advantage that
a convective envelope is not required and demands that many blue
stragglers  be in binary systems, as is observed.
During the distended contraction phase of the blue straggler's
evolution, a nearby companion could exert a considerable torque on
the star, forcing the stars to approach tidal lock.  The angular momentum
transfer to the companion will tend to force it into a larger orbit
and reduce any initial eccentricity in the orbit, as well as slowing
the rotation of the blue straggler.

\subsubsection{Surface Abundances.}

Although it is questionable whether surface convection is sufficient to
explain the moderate rotation rates of most blue stragglers, any
amount of surface convection could act to alter the surface abundances
of these stars.  Since we find that surface convection does not occur
in the most massive blue stragglers, it is possible that abundance
anomalies might be a way in which to distinguish between the formation
mechanisms, as suggested by Sills \hbox{\em et al.}\ (1997) ---  but only for the most
massive stragglers.  The convective zones in some of our
lower mass models ($M\lesssim1.40\times M_{TO}$) can penetrate to depths where
the temperature is high enough to destroy the more fragile elements,
such as lithium 
(Pritchet \& Glaspey 1991, Hobbs \& Mathieu 1991, Glaspey \hbox{\em et al.} 1994).\nocite{prit91}\nocite{gps94}\nocite{hm91}
However, although the amount of hydrodynamical mixing which occurs
during a collision appears to be small, sufficient mixing should occur in
the envelope during the merger (due to shock heating, for example --
Sec~\ref{sec:shock}) that some chemical anomalies might be
expected, even in the absence of convection.  Additionally, meridional
currents, which should occur to some extent in rapidly rotating stars,
will also mix the surface layers even if the core is not penetrated
(Tassoul \& Tassoul, 1984).

\subsection{Core Convection.}

In the normal main sequence stars of NGC 6397 and 47 Tuc, core
convection should persist for the entire main sequence lifetime of
stars with masses greater than ${\sim}{1.40M_\odot}$ and ${\sim}{1.20M_\odot}$, respectively.  Core convection does not occur at anytime 
in our unequal mass mergers, whereas core convection appears in most
of our equal mass models computed for 47 Tuc and for a
short period of time in one  equal mass merger model computed for
NGC 6397.  No core convection occurs during the distended pre-main
sequence phase, but rather starts when the model has contracted close
to its main sequence.

That no core convection occurs in our unequal mass mergers is not
surprising due to the high central density and helium abundance.  This
is in contrast to the results of Sills \hbox{\em et al.}\ (1997) who find that
unequal mass mergers typically do develop core convection.  The
simulations of Sills \hbox{\em et al.}, however, are based on models produced
using the endproducts of polytropic collisions, which, as shown earlier
in the discussion on entropy stratification, will tend to have
hydrogen-rich cores.  The increased hydrogen abundance in the core
 results in a higher central opacity, making the central layers
 convectively unstable.

The differences in the amount of convection seen between the models
for NGC 6397 and 47 Tuc are a consequence of the fact that NGC 6397
has a metallicity which is ${\sim}{15}$ times lower than that of
47 Tuc, reducing the efficiency of the  CNO cycle for masses less
than ${\sim}{1.40M_\odot}$.  At that mass our equal mass mergers for NGC 6397
($M_{TO} \sim 0.71M_\odot$) have sufficiently low central hydrogen
content that convective transport is not necessary.

\subsection{Consequences of Calculated Mixing Scales.}

According to the Vogt-Russell Theorem, the structure of a star in hydrostatic and thermal equilibrium
  is determined by its
mass and composition profile, 
which maps into a point on the H-R diagram.  We would expect that a star  perturbed
slightly from its position of equilibrium on the H-R diagram would
relax back to the same equilibrium position and structure.
If the star is
perturbed from its equilibrium state to a greater extent, such that no
additional convection occurs which might change its chemical profile
significantly, and such that no mass is lost during the perturbation, we
would still expect it to relax back to the same equilibrium state on
a timescale equal to the star's Kelvin-Helmholtz time scale. 
Similarly, two merger remnants produced in collisions involving
identical sets of stars and having identical masses and chemical
profiles, but which are initially at different points in the H-R diagram, 
should produce identical blue
stragglers if no significant mixing occurs during their pre-main
sequence phase.

This same argument was used earlier to explain why the exact details
of the mechanism used to expand our blue straggler models to their initial
pre-main sequence position are unimportant.  However, here it has the
additional implication that two theoretical merger remnant models
which have identical masses and chemical profiles, but were produced
using different assumptions about their structure,  should produce
identical blue stragglers 
if no significant convective mixing occurs
during their pre-main sequence phase.  For this reason, we expect the
evolution of the models of 
Sills \& Lombardi (1997)\nocite{sl97}
and of Sills \hbox{\em et al.}\ (1997) to be similar to the
evolution of our own
unequal mass merger models and equal mass merger models, respectively.

\subsection{Comparison With Observations.}

Shown on each of the tracks in Figures 4 and 5 are equally spaced
intervals of 0.05Gyr. 
Since the probability of seeing a star in any phase of its evolution
is proportional to the amount of time it spends in that phase, a blue
straggler which was created at some random time in the past has an
equal probability of being seen in any of the marked intervals on the
appropriate evolutionary track.  This
implies that the observed blue straggler distribution should cluster
roughly where the marked intervals are closest {\em if} the blue
stragglers have been created by processes similar to those represented
by the models.

Figures 6 and 7 show the observed blue straggler distributions in NGC
6397 
(Lauzeral \hbox{\em et al.}, 1992)\nocite{lauzeral92}
 and 47 Tuc 
(Guhathakurta \hbox{\em et al.}, 1992)\nocite{guhatha92}
compared with the {\em expected} distribution
\footnote{
  The expected distribution was found by extracting stars from the
  tracks at random, using probability distributions created from the
  evolution timescales of the tracks themselves.  The two distributions
  of blue stragglers, the observed and expected distributions, have the
  same distribution of apparent masses, excluding the extreme blue
  stragglers as noted in the text.  By this we mean that a comparison of
  the photometric positions of the blue stragglers and our tracks yields,
  for each blue straggler, a probability distribution for the mass.  The
  fake blue stragglers have masses drawn from this probability
  distribution.  The distribution of fake blue stragglers also takes
  into account the expected photometric errors as indicated on the
  figures.  The probabilities quoted in the text are from Monte Carlo
  simulations which will be described in Ouellette \& Pritchet (1998).
} 
of blue stragglers from
our equal and unequal mass mergers.  Ignoring, for the moment, those blue stragglers which lie
blueward of the ZAMS, the blue straggler distributions in both clusters
lie roughly in the region predicted by our unequal mass mergers; the
blue stragglers formed through equal mass mergers tend to cluster too
near to the ZAMS.  For our unequal mass mergers, the observed and
expected distributions are similar at a 68\% and 96\% confidence level
for NGC 6397 and 47 Tuc, respectively; the same comparison for our
equal mass mergers yields 18\% and 2\% confidence levels (see
Ouellette \& Pritchet 1998b\nocite{op98} for details).
The obvious interpretation of this is that most of the blue stragglers
in these two clusters may have been
created through unequal mass mergers or a similar event which leaves
the remnant in an apparently evolved state after formation (binary coalescence?). 

There is, however, an
additional, more critical observation to be made from the apparent agreement
between the distribution of the blue stragglers and the predictions of
our unequal mass merger models.
The collisional scenario for the formation of blue stragglers predicts
that such mergers will, on average, be born rapidly rotating.  This
rapid rotation can affect the evolution of the star by providing
non-thermal pressure support, which can extend the MS lifetime of a star,
or by initiating meridional circulation currents, which could mix the
star.  Non-thermal pressure support would extend the lifetime of a
star by requiring less energy to be liberated from the core in the
form of nuclear reactions, thereby prolonging the hydrogen-burning
lifetime of the star.  Additionally, the star will appear cooler and
less luminous than its non-rotating equivalent 
(Clement, 1994).\nocite{clement94}
Meridional circulation currents are a consequence of the distortion of
the star's gravitational potential by the rotation; as a result of this
distortion, circulation currents will be initiated and mix the star.
If hydrogen-rich 
material is brought into the core of the star by the circulation
currents, the hydrogen-burning lifetime of the star, and so its
lifetime as an observable
blue straggler, will be greatly
extended.  Also, due to the increased central hydrogen abundance,
the
position of the blue straggler on the CMD should be closer to the ZAMS
than it would be otherwise.  

If blue stragglers created by collisions are rotating rapidly enough
for these effects to be important, the agreement between the
predicted blue straggler distribution from our tracks,
which do not include any effects of rotation, and the observed blue
straggler distribution should be poor.  The apparent agreement on the
CMD between
the prediction of the unequal mass mergers and the observed blue
stragglers places limits on the importance of the effects of rotation.
First, although the hydrogen burning lifetimes of the blue stragglers might
in fact be different than what our models predict, the difference
cannot be extreme (certainly much less than a factor of two) else the observed blue
straggler distribution would be significantly more clustered near the
MS.  Second, if meridional
circulation were to mix the star, the observed blue stragglers would have
significant scatter blueward toward the ZAMS.  This does not exclude
meridional circulation in the envelope (e.g. Tassoul \& Tassoul, 1984), but
little or no helium-rich material will be brought to the surface in
such a case.

\subsubsection{Blue Straggler Formation Rates.}

A simple estimate of the formation rate of blue stragglers in these
clusters can be made by finding the average amount of time our merger
models will spend as observable blue stragglers.  Assuming, as
suggested by the comparison of the observed and expected blue
straggler distributions, that the blue stragglers in these clusters
are formed by unequal mass mergers, the average formation rates of
blue stragglers in NGC 6397 and 47 Tuc are 1 blue straggler every
${\sim}{5.0}{\times}{10^7}$yr and ${\sim}{3.6}{\times}{10^7}$yr,
respectively.  For 
comparison, if the blue   
stragglers in these clusters were formed solely by equal mass mergers
the rates would be 1 blue straggler every ${\sim}{2.3}{\times}{10^8}$yr and
${\sim}{1.2}{\times}{10^8}$yr.  Assuming that the binary fraction in the
core of each cluster is 100\% (following Leonard \& Fahlman, 1991),
and that the average binary semi-major 
axis is $\sim0.1$AU, equation $4.12$ of Sigurdsson \& Phinney
(1993)\nocite{sig93} yields an average time between collisions of
$\sim{6}{\times}{10^7}$yr and $\sim{7}{\times}{10^5}$yr for NGC 6397 and
47 Tuc, respectively.  These collision timescales are close to lower
limits, since the binary fraction is most likely not 100\%; for the
timescale for 47 Tuc to agree with the model predictions, the binary
fraction would have to be ${\sim}10\%$, or the average binary
separation would have to be smaller.
(A more 
detailed analysis will be presented in Ouellette \& Pritchet 1998b).

%==============================================================

\subsection{Extreme Blue Stragglers.}

In the above discussion and analysis we have pointedly ignored the
blue stragglers which lie blueward of the ZAMS.  While it is possible
that some of these `extremely blue' blue stragglers are merely field
objects\footnote{Guhathakurta \hbox{\em et al.}\ (1992) estimate that $\lesssim 3$ of
  the blue stragglers they observed in 47 Tuc can be explained by
  background objects in 
  the SMC.}
 or photometric errors, many other studies of clusters with
 excellent photometric quality and low foreground object
 contamination (e.g. Bolte 1992\nocite{bolt92}) exhibit such extreme blue stragglers as well.  This
 suggests that these blue stragglers are produced by a different
 process than that modelled here.  In fact, as other studies have
 shown 
(Sandquist \hbox{\em et al.}\ 1997,\nocite{sbh97}
Sills \hbox{\em et al.}\ 1997,\nocite{slbd97}
Ouellette \& Pritchet 1996),\nocite{op96}
these stars lie roughly where we would expect fully
 mixed merger remnants to be.  However, inspection of the `fake' blue straggler
distributions in Figures 6 and 7 shows that photometric scatter may be
responsible for some of these stars.

Earlier we suggested that meridional currents produced by
rapid rotation are not necessarily 
important when modelling blue stragglers produced by collisions.  It
may in fact 
be that some rotational threshold exists above which this process
will become important for the more rapidly rotating blue stragglers.

Mestel (1953)\nocite{mestel53} pointed out that, although the average rotation
rate of early type (A-F) stars is rapid enough that one would expect
meridional circulation to be ongoing within these stars, the fact that
we see such stars evolved to the giant phase implies that they have
chemically inhomogeneous structures, which would not be the case if meridional
circulation currents were present.  Mestel (and 
Tassoul \& Tassoul, 1984)\nocite{tassoul84}
 found that meridional currents were strongly inhibited
by even a slight chemical gradient throughout the star.  Since the
stars in a GC are evolved and so not chemically homogeneous,
and we expect chemical gradients in merger remnants, this may be
enough to disrupt circulation currents in the majority of blue
stragglers.  Tassoul \& Tassoul (1984) found that, for stars rotating
well below rotational break-up, the velocity of the circulation
currents increased with increasing rotational velocity, but the depth
of penetration of the currents was virtually unchanged  --- hence it
is very unlikely that meridional currents could be invoked to mix fresh
hydrogen to the cores of these stars.  The predicted
circulation in the envelope may still be enough to affect observed chemical
abundances, although it is questionable whether or not enough helium
will be mixed 
to the surface to cause the star to appear as an extreme blue
straggler.  However, it may be possible that, if a star was rotating
near to rotational break-up, the meridional currents might be able to
penetrate further into the star than in Tassoul \& Tassoul's more slowly
rotating models --- this would have to be confirmed through additional
theoretical studies into rotation using two-dimensional stellar models
and is purely speculative at this point.
If meridional currents are responsible for mixing helium to the
surface of extreme blue stragglers, then these stars are most likely
those which were created with the highest initial rotation velocities.

%==============================================================

\section{Conclusions.}

We have evolved stellar models which are appropriate for stars created
during collisions between equal and unequal mass stars.  In doing so
we have used the results of SPH simulations of colliding polytropes,
most importantly the predictions of entropy stratification.  To
produce these models, we have made several assumptions to facilitate
the incorporation of the results of the SPH simulations of collisions
between {\em polytropic} stars into models of real stars; these
assumptions include neglecting mass loss, rotation, and shock heating
during the collision.  Because we have restricted our attention to
head-on parabolic collisions, we believe these approximations to be
reasonably valid;  mass loss and rotational velocity are shown to be
small in this case by the simulations of LRS; shock heating is shown
to be small and, most importantly, should not affect the
distribution of helium-rich gas in the merger remnant
(Sec~\ref{sec:shock}), although it is likely that shock heating will
affect the distribution of material near the surface of the star.  The
form of the energy injection term $\epsilon_x$ which we use to expand
our merger remnant models is constrained by the results of SPH, but
its form is nonetheless not crucial as long as significant mixing is
not artificially introduced (via the Vogt-Russell Theorem; Vogt 1926,
Russell 1927).  In comparing the predictions of these models with the
blue stragglers seen in two dense clusters, NGC 6397 and 47 Tuc, we
have also assumed that all of the blue stragglers in these clusters
have been formed through collisions.

The apparent agreement between the predictions of our unequal mass
merger models and the observed blue straggler distributions  NGC 6397
and 47 Tuc suggests that 
\begin{itemize}
  \item little or no mixing occurred during the formation of these
    blue stragglers, either of fresh hydrogen to the cores, or of  helium
    to the surfaces.  As explained earlier, such mixing would produce
    a significant blueward scatter in the distribution of blue
    stragglers in the CMD --- as it is, our models with the best
    agreement with the observations, the unequal mass mergers, produce
    rather red stars, relative to the clusters' ZAMS.  Regardless of the
    formation mechanism for the 
    blue stragglers in these clusters, it is apparent from their
    location redward of the ZAMS that they are formed as
    {\em evolved} stars (Ouellette \& Pritchet, 1996).
  \item there may be some form of efficient AML mechanism acting to
    slow the rotation of these stars.  Blue stragglers which are formed by
    collisions are expected to be rapidly rotating: the affect of
    rapid rotation on the observed distribution of blue stragglers in
    the CMD should result in a poor agreement with our non-rotating
    models, and yet the agreement appears to be quite good.   The
    thin, short-lived convective envelopes seen in our 
    models precludes a magnetically driven stellar wind AML mechanism to explain
    the slow rotation rates and apparent lack of rotational effects; however,
    angular momentum transfer to a circumstellar disk or nearby
    companion are still plausible.

\end{itemize}

The lack of surface convection in our most massive models and the
thin convective envelopes seen in our less massive models means that
surface abundances of collisionally
merged blue stragglers might not be altered from those of the parent stars,
unless some amount of hydrodynamical mixing occurs during the
collision.  Additionally, meridional circulation currents, which
should be confined to the stellar envelope except in the most extreme
cases, may act to mix the surface layers, regardless of the amount of
convection.

The extreme blue stragglers observed in both NGC 6397 and 47 Tuc
occupy the region of the CMD which should be populated by highly mixed
stars.  It is possible that these blue stragglers are the few that
were born rotating rapidly enough that  the meridional circulation
currents are able to penetrate more deeply in to the star; if so,
these stars 
should still be rapidly rotating.  It is, however, possible that
some fraction of these stars are merely
photometric errors.

%==============================================================

\section{Acknowledgements}

We would like to thank both Russell Robb and Don VandenBerg for
providing useful comments on an earlier version of this manuscript.
Additional thanks must go to Dr.\ VandenBerg for allowing us to use his
stellar evolution code, and for numerous insightful discussions on the
topic of stellar evolution. Finally, we wish to thank the Natural
Sciences and Engineering Research 
Council for financial support through a Research Grant to C.J.P..

%==============================================================

\newpage

\bibliographystyle{astron}

%==============================================================

\newpage

\figcaption[Ouellette.Fig1.ps] {Density ($\rho$) and entropy ($S$) profiles for
  representative polytropes.  Shown are profiles for  0.80M$_\odot$
  (solid line, $n=3$, $R=1.0R_\odot$),
  0.60M$_\odot$ (dashed line, $n=3$, $R=0.56R_\odot$), and
  0.40M$_\odot$ (dotted line, $n=3/2$, $R=0.37R_\odot$)
  polytropes, calculated using the radii and 
  polytropic indices given.  The 0.6M$_\odot$ composite polytrope has
  an $n=3$ core and an $n=1.5$ envelope with the boundary between the
  two located at $R_{bnd}=0.29R_\odot$. \label{Figure 1.}}

\figcaption[Ouellette.Fig2.ps] {Density ($\rho$) and entropy ($S$) profiles for real GC stars.    Shown are profiles for 0.864M$_\odot$ (solid line),
0.60M$_\odot$ (dashed line), and 0.40M$_\odot$ (dotted line) stellar
models at an age (14 Gyr) and metallicity ([Fe/H]=-0.71) appropriate
for 47 Tuc. \label{Figure 2.}}

\figcaption[Ouellette.Fig3.ps] {Comparison of pre-shock (thin lines)
  and post-shock (thick lines) entropy profiles for the stars shown in
  Figure 2.  See Section~\ref{sec:shock} for a description of the
  calculation of the amount of entropy production during shock
  heating} \label{Figure 3.} 

\figcaption[Ouellette.Fig4.ps]{ Evolutionary tracks (thin solid lines) for equal mass (top) and unequal mass (bottom)
  merger models for NGC 6397. Masses of the models range from
  0.90M$_\odot$ to 1.40M$_\odot$ in steps of 0.10M$_\odot$.  The open
  circles ($\circ$) are placed along the tracks at equal intervals of
  0.05Gyr.  Also shown are the theoretical cluster ZAMS (thick solid
  line) and a 19 Gyr isochrone (thick dashed line).  Note the lack of
  a convective ``hook'' in all of the models, except for the most
  massive equal mass merger model. \label{Figure 4.}}

\figcaption[Ouellette.Fig5.ps] {Similar to Figure 4, except for 47
  Tuc.  The masses of the models range from 1.10M$_\odot$ to
  1.70M$_\odot$ in steps of 0.10M$_\odot$. Also shown
  is a 14 Gyr isochrone. Symbols and line styles have the same
  meanings as in Figure 5. \label{Figure 5.}}

\figcaption[Ouellette.Fig6.ps] {The observed distribution of blue stragglers in NGC
  6397 ($\bullet$) compared with blue stragglers created from the tracks of our
merger models ($\Box$, see text).  Also shown are the
tracks for the appropriate mergers with the `pre-blue straggler' phase
omitted for clarity (thin dotted lines).  The error bar at the top of
the figure 
corresponds to the random photometric errors appropriate for the observations.
The theoretical cluster ZAMS (thick solid line) and a 19Gyr isochrone
(thick dashed line) are also shown. \label{Figure 6.}}

\figcaption[Ouellette.Fig7.ps]{Similar to Figure 6, except for 47 Tuc.  Also shown
  is a 14 Gyr isochrone. Symbols and line styles have the same
  meanings as in Figure 6. \label{Figure 7.}}

\newpage
\begin{table}[ht]

\begin{center}

\begin{tabular}{lcc}
\multicolumn{3}{c}{Table 1: Cluster Parameters}\\
\tableline\tableline
Parameter & NGC 6397 & 47 Tuc \\
\tableline
 \multicolumn{3}{c}{Accepted Parameters}\\
{[}Fe/H{]} & -1.91 & -0.71 \\
{[}$\alpha$/Fe{]} & 0.30 & 0.30 \\
$E(B-V)$ & 0.18 & 0.04 \\
V(HB) & 12.90 & 14.09 \\
c\tablenotemark{a} & 2.50 & 2.04 \\
 \multicolumn{3}{c}{Derived Parameters}\\
$M^{HB}_V$ & 0.598 & 0.838 \\
$(m-M)_V$ & 12.30 & 13.25 \\
Age(Gyr) & 19.0 & 14.0 \\
$M_{TO} (M_\odot)$ & 0.708 & 0.864 \\

\tableline

\end{tabular}

\end{center}

\tablenotetext{a}{
Central concentration. Defined as $c=\log(r_{tidal}/r_{core})$.
}
\end{table}

\newpage

\begin{figure}
\centerline{
\epsfbox{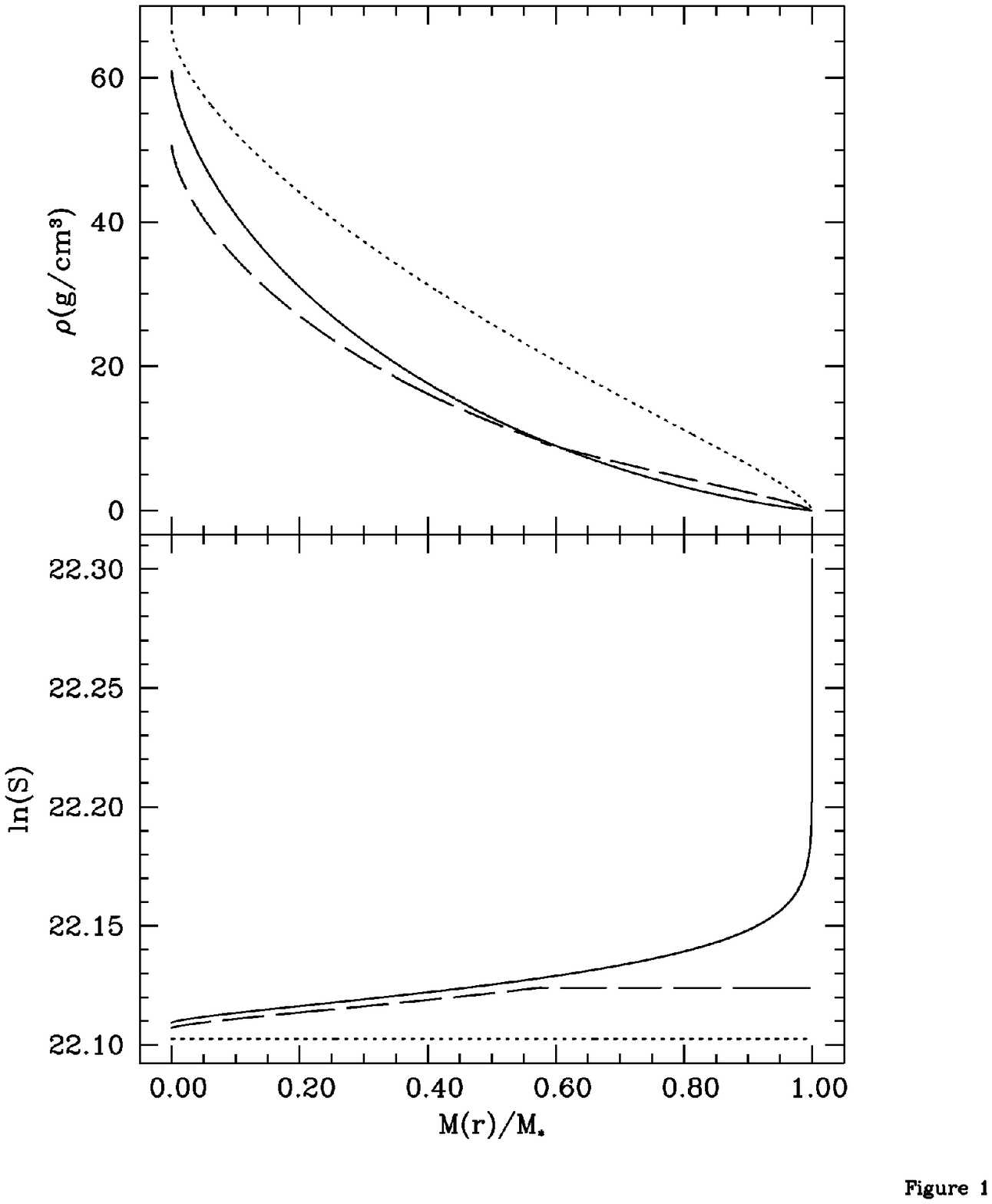}
}
\end{figure}

\newpage

\begin{figure}
\centerline{
\epsfbox{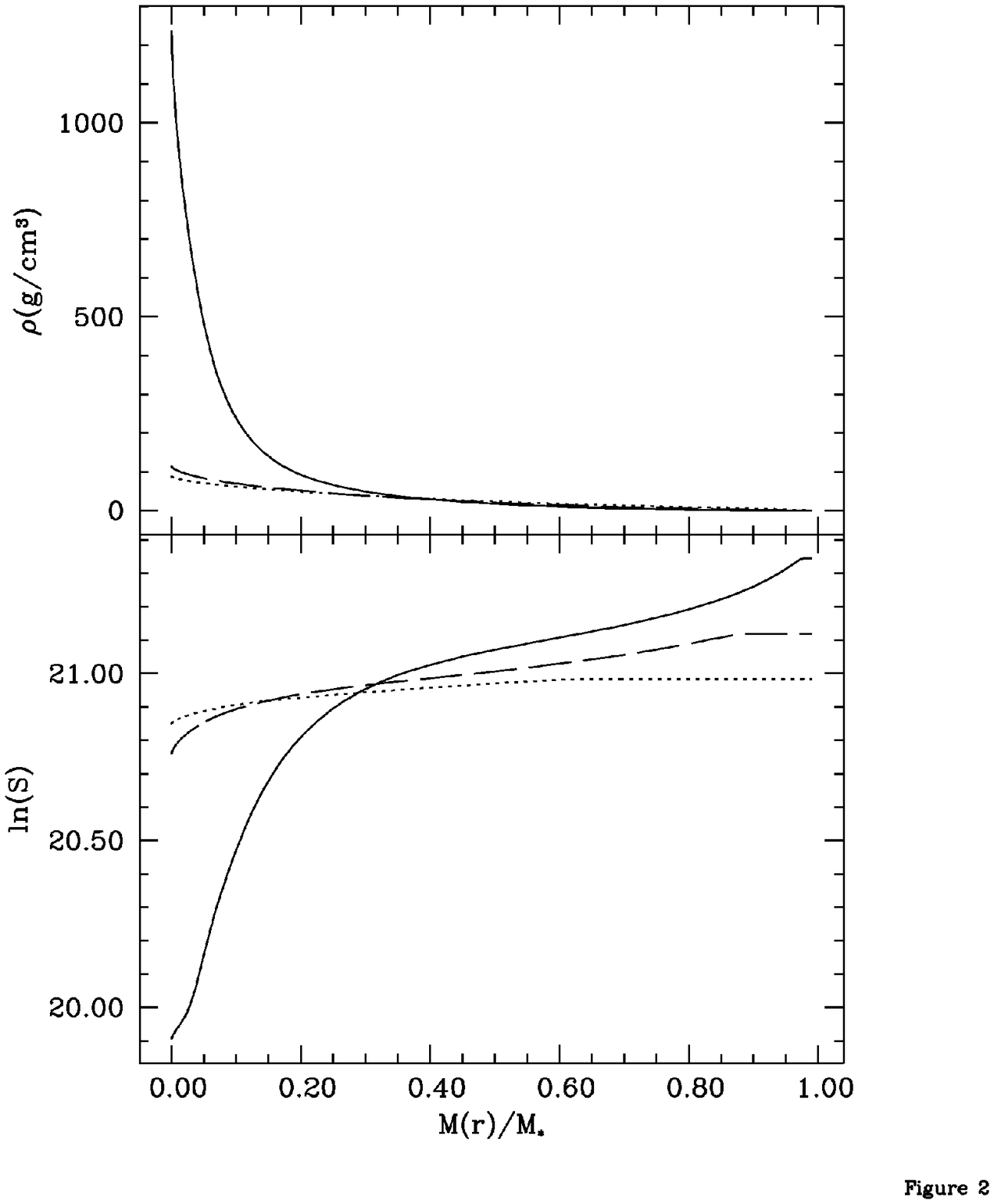}
}
\end{figure}

\newpage

\begin{figure}
\centerline{
\epsfbox{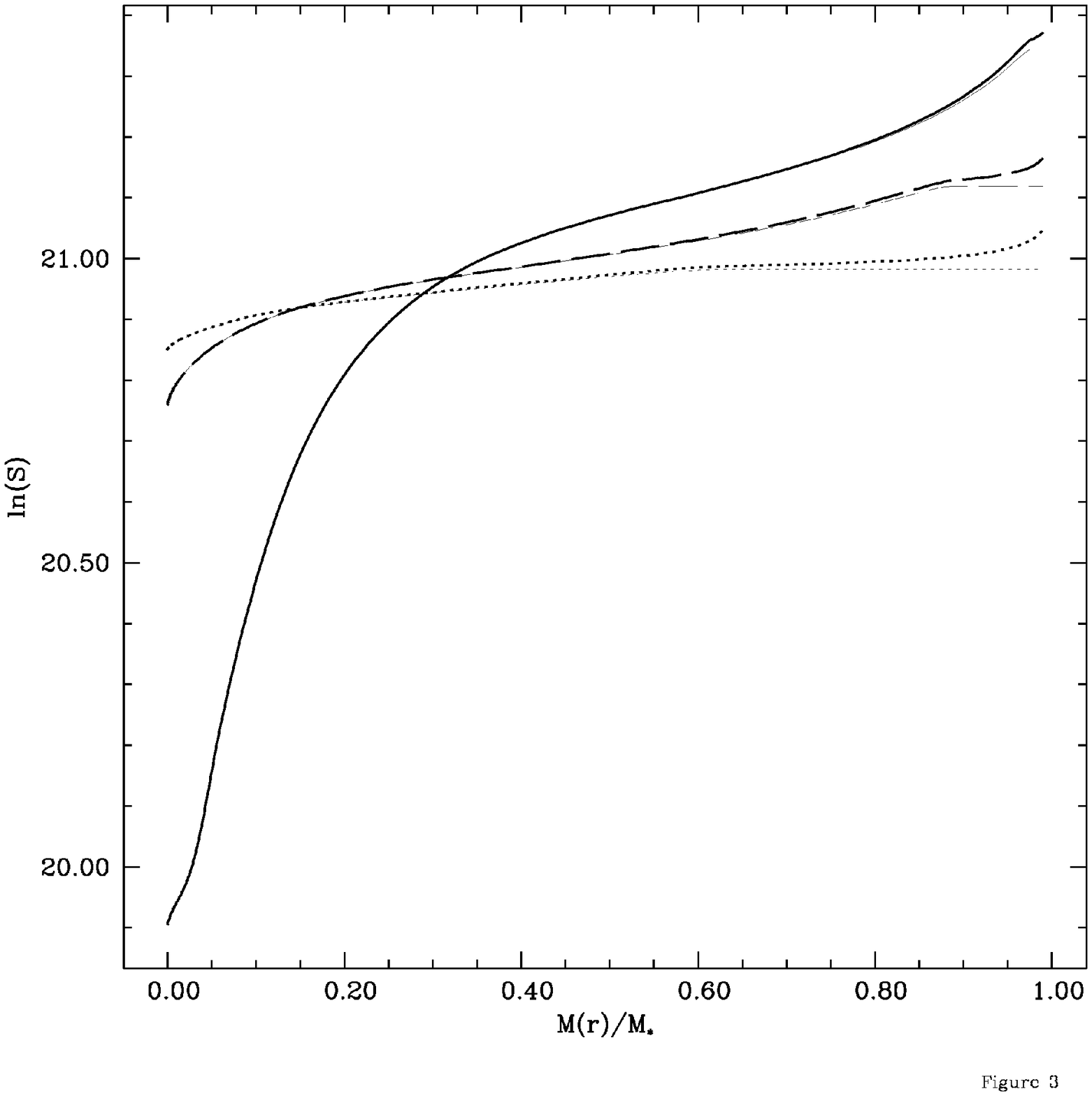}
}
\end{figure}

\newpage

\begin{figure}
\centerline{
\epsfbox{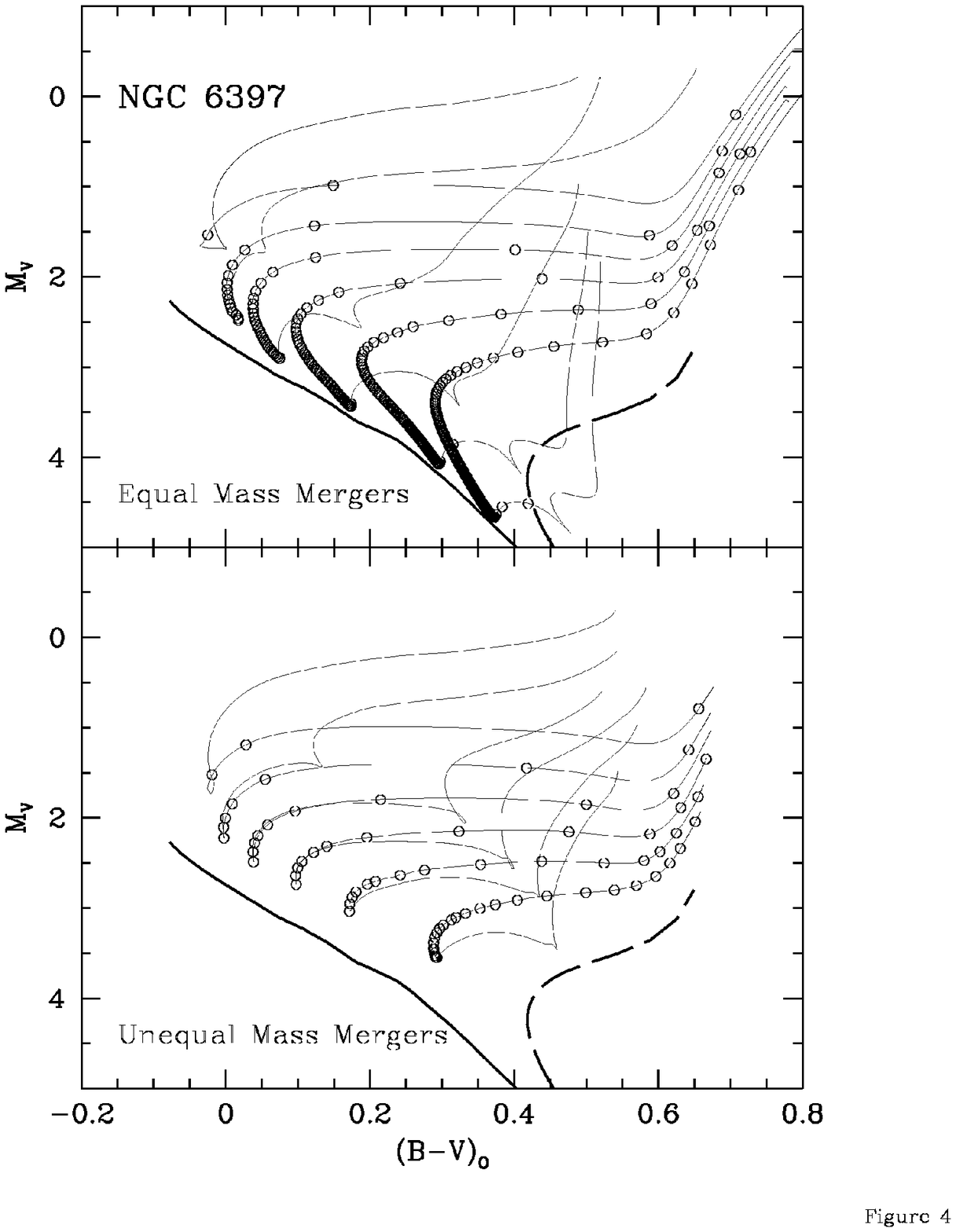}
}
\end{figure}

\newpage

\begin{figure}
\centerline{
\epsfbox{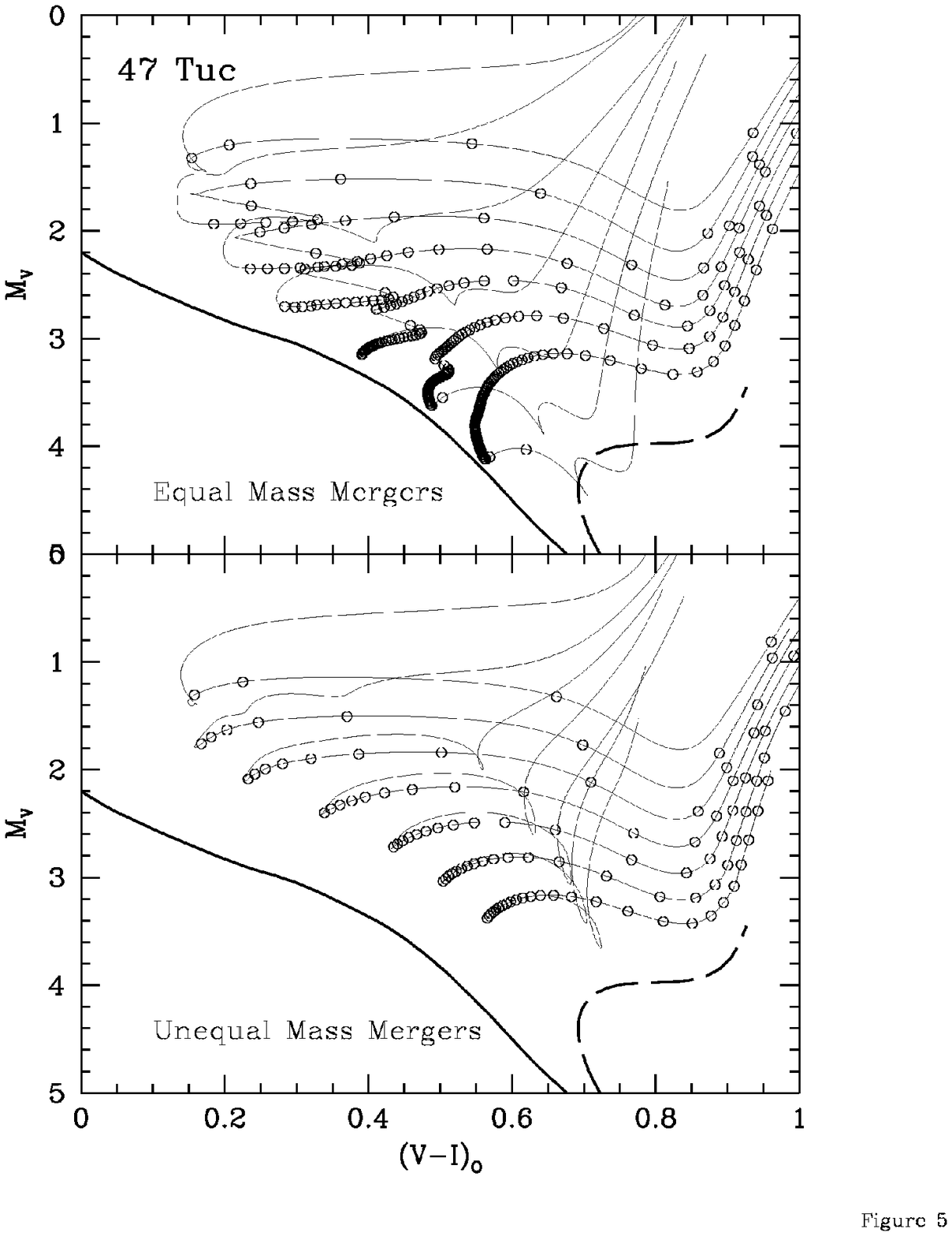}
}
\end{figure}

\newpage

\begin{figure}
\centerline{
\epsfbox{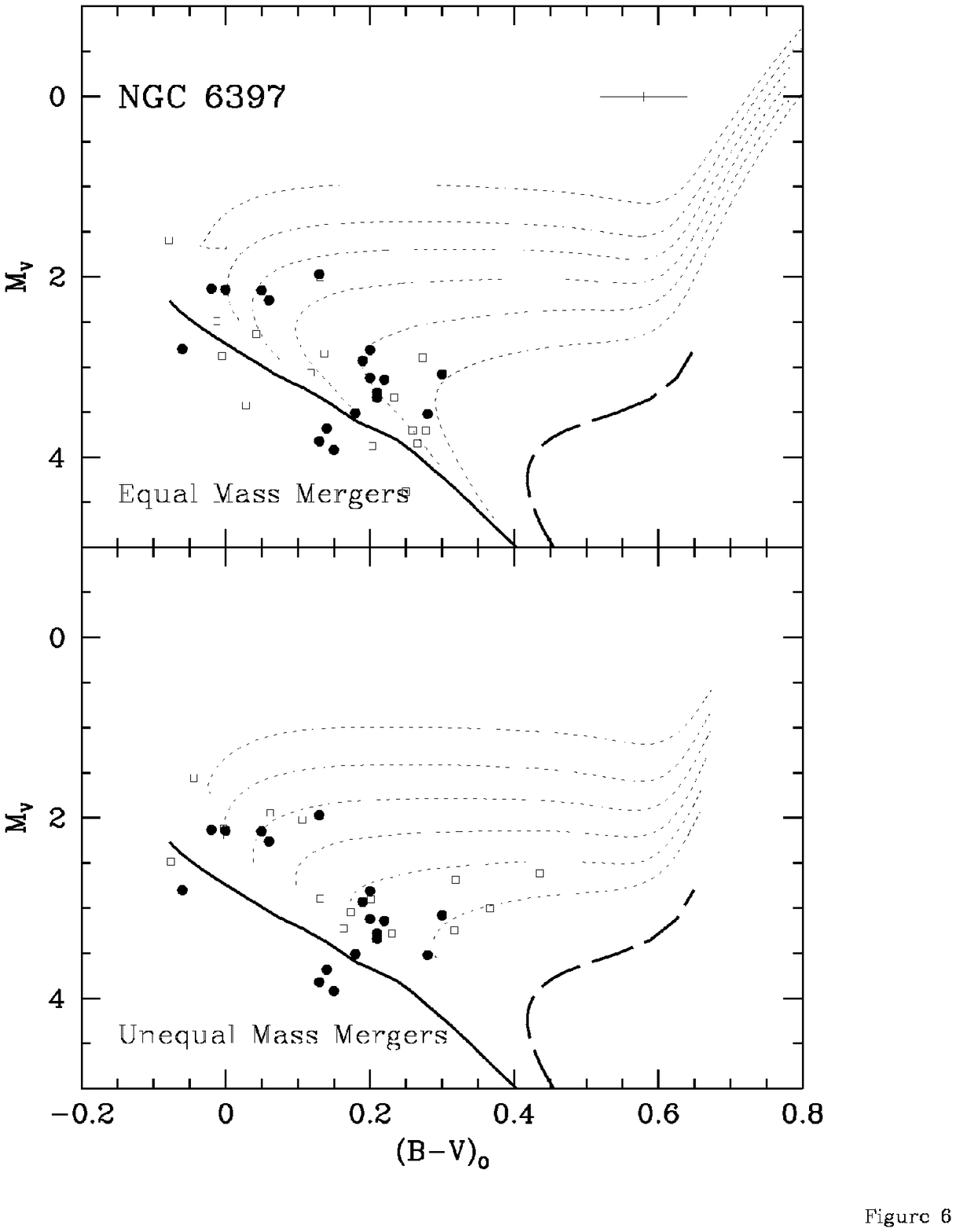}
}
\end{figure}
\newpage

\begin{figure}
\centerline{
\epsfbox{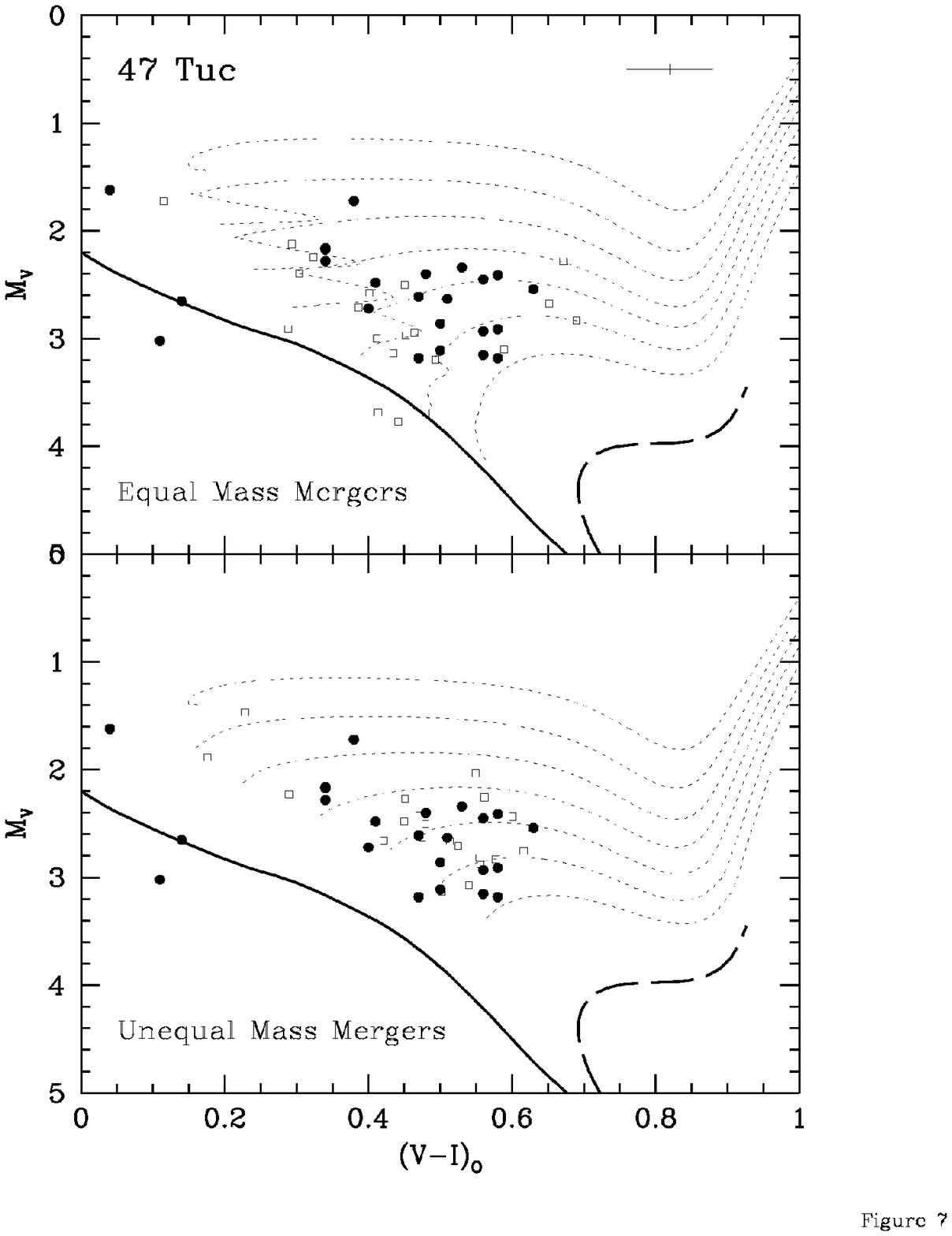}
}
\end{figure}
\end{document}